# Designing a parallel suffix sort

Kunal Chowdhury

**Abstract**— Suffix sort plays a critical role in various computational algorithms including genomics as well as in frequently used day to day software applications. The sorting algorithm becomes tricky when we have lot of repeated characters in the string for a given radix. Various innovative implementations are available in this area e.g., Manber Myers. We present here an analysis that uses a concept around generalized polynomial factorization to sort these suffixes. The initial generation of these substring specific polynomial can be efficiently done using parallel threads and shared memory. The set of distinct factors and their order are known beforehand, and this helps us to sort the polynomials (equivalent of strings) accordingly.

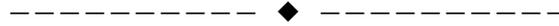

## 1 INTRODUCTION

A sequence of characters is a string, and it forms an important fundamental abstraction in genomic sequences, information processing systems, communication and programming systems. Typically, they are 8-bit or 16-bit integers in C and Java respectively. As string comprises of characters, string sort has received a lot of attention, and quite a few important algorithms have been discovered in this regard.

Typically, any compare-based algorithm requires $NlgN$ compares, however radix-based sorting has been able to bring down this running time significantly e.g., LSD sort has a guaranteed running time of $2W(N + R)$ where W is the fixed length of keys, R is the radix and N denotes the number of strings. Similarly, MSD too is a radix-based sort and guaranteed time complexity for a random input is $Nlog_R N$. A related problem in string sort is suffix sort where the worst time complexity is typically $N^2 logN$. However, one of the key algorithms viz, Manber-Myers has linearithmic time complexity.

Our effort in this discussion is to consider strings as polynomials defined using a non-commutative operator and then come up with a solution to design the sorting algorithm in a multiprocessor environment /GPU. This kind of sort harnesses the power of parallel thread processing and cache coherency leading to finer control of the end-to-end process. One can also add custom logic in these threads to perform any further business functionality.

• *Kunal Chowdhury. E-mail:2019ht12471@ wilp.bits-pilani.ac.in*

## 2 STRINGS AND POLYNOMIALS

A string $s = c_1 c_2 \ldots c_{n-1} c_n$ containing the characters $c_1, c_2, \ldots c_{n-1}, c_n$ can be represented as a polynomial $P(n) = x^0. \ x^1. x^2. \ldots . x^{n-1}$ on $x$ using an operator dot(.) that is equivalent to concatenation. However, we can have cases where there is repetition of characters in the string and in such a case, we can count the consecutive instances of the given character and use it as a coefficient in the polynomial. Thus, we can define the polynomial as $P(n) = c_0 1. c_1 x^1. c_2 x^2. \ldots . c_{n-1} x^{n-1}$ where $c_0, \ c_1, \ \ldots$ etc. are the corresponding coefficients of $x$.

With this notion two strings $s$ and $t$ with corresponding polynomial notations $P(n)$ and $Q(n)$ respectively can be compared as follows

Let $a$ and $b$ denote the smallest degrees of $x$ in $P(n)$ and $Q(n)$ respectively where one of the following conditions is satisfied:

1. Either $a$ and $b$ are different

2. The corresponding coefficients $c_a$ and $c_b$ are different.

We now define the ordering of $P(n)$ and $Q(n)$ as follows:

1. If $a < b$ , then $P(n) < Q(n)$ else $P(n) > Q(n)$

2. If $c_a < c_b$ , then $P(n) < Q(n)$ else $P(n) > Q(n)$

### 2.1 Polynomial Factorization

We define the factors of a polynomial denoting the string $s = c_1 c_2 \ldots c_{n-1} c_n$ by grouping the consecutive characters using the following rules:

1. The characters are monotonically increasing i.e., for some $p$ and $q$ (with $p < q$) we have $c_p < c_{p+1} < \cdots < c_q$

2. The characters are same i.e., for some $p$ and $q$ (with $p < q$) we have $c_p = c_{p+1} = \cdots = c_q$

Now we present the factorization process for the above conditions

1. As the characters are different, we group them into a smaller polynomial as $x^p . x^{p+1} . \ldots . x^q$
2. We attach the coefficient $q - p$ to $x^p$

It is trivial to verify that the above notion of polynomial ordering and factorization lies in line with the concept of lexical sorting of strings.

## 3  SUBSTRING SORT

The $n$ different substrings of $s = c_1 c_2 \ldots c_{n-1} c_n$ will have a fixed set of factors corresponding to the strings $c_1 c_2 \ldots c_{n-1} c_n$ , $c_2 \ldots c_{n-1} c_n$ , $c_3 \ldots c_{n-1} c_n$ … till $c_n$ . If all the characters are different, then we have $n$ distinct factors whereas for the case when all the characters are same, we have distinct coefficients for the same factor.

Once the factors are precomputed, we order them into buckets $B_1, B_2 , \ldots B_p$ for $p$ distinct factors as per the ordering of the corresponding polynomial.

Each bucket holds the residue and coefficient obtained after dividing the polynomial (for the corresponding substring) by a given factor.

If X and Y are two polynomials as defined above and the string, $s = X . Y$ then it's easy to verify that residue when $s$ (as a polynomial) is divided by X will be Y. However, any other polynomial Z will not divide s., It should be noted that this operation is not commutative, hence $s = X . Y \neq Y . X$

### 3.1  Proposition

*If the substrings are divided into buckets (specific to the precomputed factors) containing the pair of residue and coefficients obtained after dividing the substring polynomial by the factor, then the ordered set of buckets hold the sorted substrings.*

**Proof**: The polynomial representation of the strings is well ordered according to the degree and coefficient of their factors defined by the concatenation operator. In conventional notation this can be regarded as a non-commutative + operator. As each bucket is sorted as per this principle and the buckets themselves are sorted according to the polynomial representation of the factor, the result follows.

## 4  EXAMPLES

Following represent some concrete examples based on the concept we discussed so far.

### 4.1  String -babaabcbabaa

| Character | Representation |
|-----------|----------------|
| a | 1 |
| b | $x$ |
| c | $x^2$ |

Let us write down all the substrings of this string and their corresponding polynomial representation as shown in Table 1.1

Now we sort the buckets and residues within each bucket as shown below as 1.2

It is easy to verify that the residues of buckets $x$ use the already sorted residues of buckets 1 and 1.x respectively thus reducing the extra computation.

### 4.2  String – twinstwins

| Character | Representation |
|-----------|----------------|
| i | 1 |
| n | $x$ |
| s | $x^2$ |
| t | $x^3$ |
| w | $x^4$ |

Corresponding substrings and their buckets are shown in Table 2.1 and the sorted buckets are shown in Table 2.2

### 4.3  String – aacaagtttacaagc

| Character | Representation |
|-----------|----------------|
| a | 1 |
| c | $x$ |
| g | $x^2$ |
| t | $x^3$ |

Corresponding substrings and their buckets are shown in Table 3.1 and the sorted buckets are shown in Table 3.2

| String | Polynomial Representation | Bucket Assigned |
|---|---|---|
| babaabcbabaa | $x.(1.x).2.(x.x^2).x(1.x).2$ | $x$ |
| abaabcbabaa | $(1.x).2.(x.x^2).x.(1.x).2$ | $1.x$ |
| baabcbabaa | $x.2.(x.x^2).x.(1.x).2$ | $x$ |
| aabcbabaa | $2.(x.x^2).x.(1.x).2$ | $1$ |
| abcbabaa | $(1.x.x^2).x.(1.x).2$ | $1.x.x^2$ |
| bcbabaa | $(x.x^2).x.(1.x).2$ | $x.x^2$ |
| cbabaa | $x^2.x.(1.x).2$ | $x^2$ |
| babaa | $x.(1.x).2$ | $x$ |
| abaa | $(1.x).2$ | $1.x$ |
| baa | $x.2$ | $x$ |
| aa | $2$ | $1$ |
| a | $1$ | $1$ |

**Table-1.1**

| Buckets | Residue | Sorted substrings |
|---|---|---|
| 1 | 1 | a |
| 1 | 2 | aa |
| 1 | $2.(x.x^2).x.(1.x).2$ | aabcbabaa |
| $1.x$ | 2 | abaa |
| $1.x$ | $2.(x.x^2).x.(1.x).2$ | abaabcbabaa |
| $1.x.x^2$ | $x.(1.x).2$ | abcbabaa |
| $x$ | 2 | *baa* |
| $x$ | $2.(x.x^2).x.(1.x).2$ | *baabcbabaa* |
| $x$ | $(1.x).2$ | *babaa* |
| $x$ | $(1.x).2.(x.x^2).x(1.x).2$ | *babaabcbabaa* |
| $x.x^2$ | $x.(1.x).2$ | bcbabaa |
| $x^2$ | $x.(1.x).2$ | cbabaa |

| $x$ | 2 | baa |
|---|---|---|
| $x$ | $2.(x.x^2).x.(1.x).2$ | baabcbabaa |
| $x$ | $(1.x).2$ | babaa |
| $x$ | $(1.x).2.(x.x^2).x(1.x).2$ | babaabcbabaa |

**Table-1.2**

| String | Polynomial Representation | Bucket Assigned |
|--------|--------------------------|-----------------|
| twinstwins | $(x^3.x^4).(1.x.x^2.x^3.x^4).(1.x.x^2)$ | $x^3.x^4$ |
| winstwins | $x^4.(1.x.x^2.x^3.x^4).(1.x.x^2)$ | $x^4$ |
| instwins | $(1.x.x^2.x^3.x^4).(1.x.x^2)$ | $1.x.x^2.x^3.x^4$ |
| nstwins | $(x.x^2.x^3.x^4).(1.x.x^2)$ | $x.x^2.x^3.x^4$ |
| stwins | $(x^2.x^3.x^4).(1.x.x^2)$ | $x^2.x^3.x^4$ |
| twins | $(x^3.x^4).(1.x.x^2)$ | $x^3.x^4$ |
| wins | $x^4.(1.x.x^2)$ | $x^4$ |
| ins | $1.x.x^2$ | $1$ |
| ns | $x.x^2$ | $x$ |
| s | $x^2$ | $x^2$ |

**Table-2.1**

| Buckets | Residue | Sorted substrings |
|---------|---------|-------------------|
| $1$ | $x.x^2$ | ins |
| $1.x.x^2.x^3.x^4$ | $1.x.x^2$ | instwins |
| $x$ | $x^2$ | ns |
| $x.x^2.x^3.x^4$ | $1.x.x^2$ | nstwins |
| $x^2$ | $1$ | s |
| $x^2.x^3.x^4$ | $1.x.x^2$ | stwins |
| $x^3.x^4$ | $1.x.x^2$ | twins |
| $x^3.x^4$ | $(1.x.x^2.x^3.x^4).(1.x.x^2)$ | twinstwins |
| $x^4$ | $1.x.x^2$ | wins |
| $x^4$ | $(1.x.x^2.x^3.x^4).(1.x.x^2)$ | winstwins |

**Table-2.2**

| String | Polynomial Representation | Bucket Assigned |
|---|---|---|
| aacaagtttacaagc | $2.x.2.(x^2.x^3).2x^3.(1.x).2.x^2.x$ | $1$ |
| acaagtttacaagc | $(1.x).2.(x^2.x^3).2x^3.(1.x).2.x^2.x$ | $1.x$ |
| caagtttacaagc | $x.2.(x^2.x^3).2x^3.(1.x).2.x^2.x$ | $x$ |
| aagtttacaagc | $2.(x^2.x^3).2x^3.(1.x).2.x^2.x$ | $1$ |
| agtttacaagc | $(1.x^2.x^3).2x^3.(1.x).2.x^2.x$ | $1.x^2.x^3$ |
| gtttacaagc | $(x^2.x^3).2x^3.(1.x).2.x^2.x$ | $x^2.x^3$ |
| tttacaagc | $3x^3.(1.x).2.x^2.x$ | $x^3$ |
| ttacaagc | $2x^3.(1.x).2.x^2.x$ | $x^3$ |
| tacaagc | $x^3.(1.x).2.x^2.x$ | $x^3$ |
| acaagc | $(1.x).2.x^2.x$ | $1.x$ |
| caagc | $x.2.x^2.x$ | $x$ |
| aagc | $2.x^2.x$ | $1$ |
| agc | $(1.x^2).x$ | $1.x^2$ |
| gc | $x^2.x$ | $x^2$ |
| c | $x$ | $x$ |

**Table-3.1**

| Row | Buckets | Residue | Sorted substrings |
|---|---|---|---|
| 1 | $1$ | $2.x.2.(x^2.x^3).2x^3.(1.x).2.x^2.x$ | aacaagtttacaagc |
| 2 | $1$ | $2.x^2.x$ | aagc |
| 3 | $1$ | $2.(x^2.x^3).2x^3.(1.x).2.x^2.x$ | aagtttacaagc |
| 4 | $1.x$ | $2.x^2.x$ | acaagc |
| 5 | $1.x$ | $2.(x^2.x^3).2x^3.(1.x).2.x^2.x$ | acaagtttacaagc |
| 6 | $1.x^2$ | $x$ | agc |
| 7 | $1.x^2.x^3$ | $2x^3.(1.x).2.x^2.x$ | agtttacaagc |
| 8 | $x$ | $1$ | c |
| 9 | $x$ | $2.x^2.x$ | caagc |
| 10 | $x$ | $2.(x^2.x^3).2x^3.(1.x).2.x^2.x$ | caagtttacaagc |
| 11 | $x^2$ | $x$ | gc |
| 12 | $x^2.x^3$ | $2x^3.(1.x).2.x^2.x$ | gtttacaagc |
| 13 | $x^3$ | $(1.x).2.x^2.x$ | tacaagc |
| 14 | $x^3$ | $2(1.x).2.x^2.x$ | ttacaagc |
| 15 | $x^3$ | $3(1.x).2.x^2.x$ | tttacaagc |

**Table-3.2**

# 5 ALGORITHM TO GENERATE POLYNOMIALS FROM STRING

1. Assign the distinct sorted characters of the string – the terms $1, x, x^2, x^3$ etc.
2. Initialize two stacks - primary and secondary and a queue.
3. Iterate the characters of the string from left to right and push the corresponding term in the primary stack
4. If the top of the stack and the current character is same, then add i.e., increment the coefficient
   a. Else if top of the stack is less than the current character, we push into the stack
   b. Else pop from the stack till it is empty into a secondary stack
      i. Pop from the secondary stack and concatenate the terms using '.' Operator
      ii. Enqueue the resultant term into a queue

Add the current character into the primary stack

# 6 REPRESENTING POLYNOMIALS AS BUFFERS

Given the fact that we are dealing with only lowercase English alphabet the factors can be represented as 32-bit integer (maximum) with the most significant 6 bits reserved for the coefficient values.

## 6.1 Individual factors

Factors will be represented as integers; the nth bit will denote the presence / absence of the term - $x^{n-1}$ as discussed above.

## 6.2 Polynomial data structure

Polynomials can be represented as byte buffer as shown below where the $0^{th}$ element denotes the size followed by the int values representing the factors.

| Size = n | Int value of the 1st factor | Int value of the 2nd factor | Int value of the 3rd factor | | | | Int value of the n-th factor |
|---|---|---|---|---|---|---|---|
| | | | | | | | |

## 6.3 Polynomial Division

The polynomial division can be implemented as an XOR operation of the first element of the polynomial byte buffer with the factor that denotes the bucket value.

A perfect division will result in either 0 or the most significant 6 bits populated.

## 6.4 Algorithm to compare two polynomials

This can be achieved in two steps -

1. We first find out the minimum of the indices where the byte buffers have different values.

This can be done efficiently using parallel threads and atomic instructions in CPU or using *atomicMin ()* instruction in GPU to avoid a linear scan.

2. Once the index has been found out we do an XOR of the corresponding values of the two-byte buffers.

3. Find out the rightmost set bit in the result $n$ using $n \& \sim(n - 1)$

4. If this bit position is not present in the first string, then it is the larger of the two else second.

## 6.5 Example

Let us compare two polynomials
$2. x. 2. (x^2. x^3). 2x^3. (1. x). 2. x^2. x$ and
$2. (x^2. x^3). 2x^3. (1. x). 2. x^2. x$

Byte Buffer 1

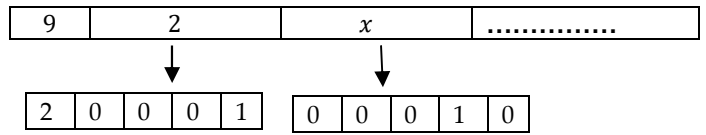

Byte Buffer 2

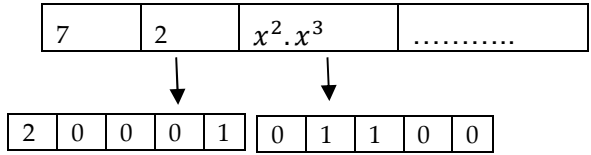

| 7 | 2 | $x^2.x^3$ | ........... |
|---|---|---|---|

| 2 | 0 | 0 | 0 | 1 | 0 | 1 | 1 | 0 | 0 |
|---|---|---|---|---|---|---|---|---|---|

*Here we have omitted the remaining elements from both the byte buffers for better visibility.*

First difference occurs between $x$ and $x^2.x^3$

We now perform XOR –

| 0 | 0 | 0 | 1 | 0 |
|---|---|---|---|---|
| 0 | 1 | 1 | 0 | 0 |
| 0 | 1 | 1 | 1 | 0 |

First set bit is at 2nd place, and is present in $x$, hence $x$ is smaller than $x^2.x^3$.

$2.x.2.(x^2.x^3).2x^3.(1.x).2.x^2.x$ is smaller than $2.(x^2.x^3).2x^3.(1.x).2.x^2.x$

## 7  FINAL DESIGN

The following flowchart represents the final design of the suffix sort using parallel threads and shared memory e.g., using GPUs.

It's worth noting that we can tune this process further in a GPU by changing the number of blocks and thread in each block.

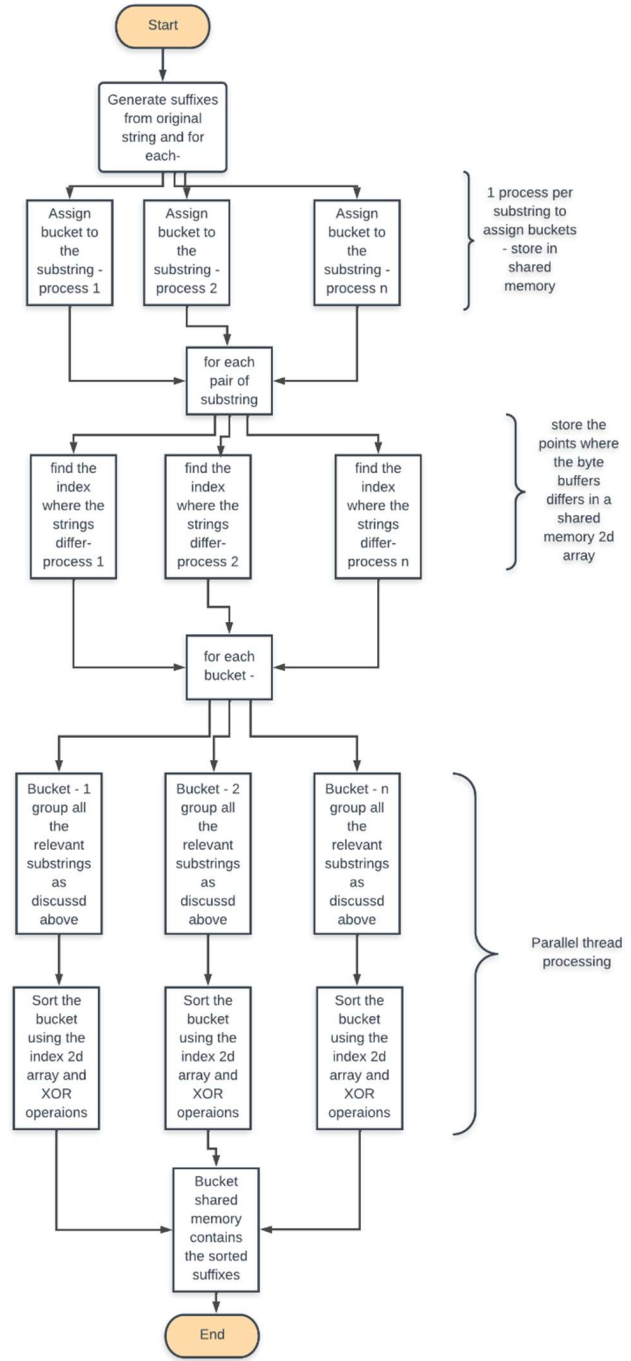

## 8  CONCLUSION

Here we discussed a strategy to sort the suffixes of a given string efficiently in a multiprocessor environment. We have investigated the polynomial representation of the strings and divided them into sorted buckets – in turn containing ordered polynomials. We also understood an implementation strategy to transform the polynomials into byte buffers and into 32-bit integer values. Future work can implement multiprocessor or GPU based sorting of these strings as libraries.